\title{Spectrum of the Anomalous Microwave Emission in the North Celestial Pole with {\it WMAP} 7-yr data}
\author{
	Anna Bonaldi\\
	Jodrell Bank Centre for Astrophysics\\
	Alan Turing Building\\
	School of Physics and Astronomy\\
	The University of Manchester\\
	Oxford Road, Manchester\\
	M13 9PL, U.K.\\
	    \and
	Sara Ricciardi\\
        INAF/IASF Bologna\\
        Via Gobetti 101\\
        Bologna, Italy\\
}

 \date{December 20 2012}

\documentclass[letterpaper,12pt, ]{article}
\usepackage{epsfig}
\begin{document}

\maketitle

\begin{abstract}
We estimate the frequency spectrum of the diffuse anomalous microwave emission (AME) on the North Celestial Pole (NCP) region of the sky with the Correlated Component Analysis (CCA) component separation method applied to {\it WMAP} 7-yr data. The NCP is a suitable region for this analysis because the AME is weakly contaminated by synchrotron and free-free emission. By modeling the AME component as a peaked spectrum we estimate the peak frequency to be $21.7\pm0.8$\,GHz, in agreement with previous analyses which favored $\nu_{\rm p}<23$\,GHz. The ability of our method to correctly recover the position of the peak is verified through simulations.
We compare the estimated AME spectrum with theoretical spinning dust models to constrain the hydrogen density $n_{\rm H}$. The best results are obtained with densities around 0.2--0.3\,cm$^{-3}$, typical of warm ionised medium (WIM) to warm neutral medium (WNM) conditions. The degeneracy with the gas temperature prevents an accurate determination of $n_{\rm H}$, especially for low hydrogen ionization fractions, where densities of a few cm$^{-3}$ are also allowed. 
\end{abstract}
\section{Introduction}

The anomalous microwave emission (AME) component is highly correlated with the far infra-red dust emission (\cite{kogut1996}, \cite{leitch1997}, \cite{banday2003}, \cite{lagache2003}, \cite{oliveira2004}, \cite{fink2004}, \cite{davies2006}, \cite{dobler2008}, \cite{ysard2010}, \cite{gold2011}) and is believed to be the electric dipole radiation from small spinning dust grains (\cite{DL1998}).  Spinning dust models predict a peaked spectrum ranging 10--150\,GHz depending on the local physical conditions (\cite{spdust1},\cite{ysard-ves2010}).  Probing the peak of the emission enables us to compare models with observations and is the best way to distinguish the AME from synchrotron emission and free-free emission, both having a power-law spectrum. 

The AME has been studied in individual dust clouds associated with reflection nebulae, molecular clouds, photo-dissociation regions and HII regions (\cite{genovasantos2011},\cite{planck2011-7.2}). 
The best examples of peaked AME spectra are probably the Perseus and $\rho$ Ophiuchi molecular clouds, which have been probed with high accuracy thanks to the availability of many different datasets.  In these regions the derived peak is at $\sim$30\,GHz (\cite{planck2011-7.2}).

The study of the AME in more diffuse regions essentially becomes a component separation problem, which is often complicated by the lack of low-frequency data covering large sky areas. Diffuse dust-correlated emission has been detected with {\it COBE-DMR}(\cite{banday2003}) and {\it WMAP} data (\cite{davies2006}, \cite{bonaldi2007}, \cite{dobler2008}, \cite{mamd}, \cite{ghosh2011}), however the details of the spectrum are still unclear. Template-fitting analyses show that the dust-correlated emission between 20 and 60\,GHz is well described by a power-law (\cite{banday2003}, \cite{davies2006}, \cite{dobler2008}, \cite{ghosh2011}). Given the error bars, a peaked spectrum is not ruled out, but a low peak frequency ($<23$\,GHz) is favored. Alternatively, the observed spectrum could result from the superposition of multiple peaked components. The component separation analysis done by \cite{bonaldi2007} found that a peaked AME model gives better results in terms of CMB cleaning, however no estimation of the AME peak frequency was attempted. 

In this work we address the estimation of the peak frequency of the diffuse AME on {\it WMAP} 7-yr data complemented by ancillary data with the Correlated Component Analysis (CCA, \cite{ricciardi2010}) component separation method.
We consider the North Celestial Pole (NCP) region of the sky, centered on Galactic coordinates $(l,b)\sim(125^\circ,25^\circ)$. As already noted by \cite{davies2006}, this region of the sky is particularly suited for this analysis since there is significant thermal and anomalous dust emission, while synchrotron emission and free-free emission (traced by the 408 MHz from \cite{haslam} and the H$\alpha$ map from \cite{cliveff}) are faint. 


\begin{figure}
\begin{center}
\includegraphics[width=6cm]{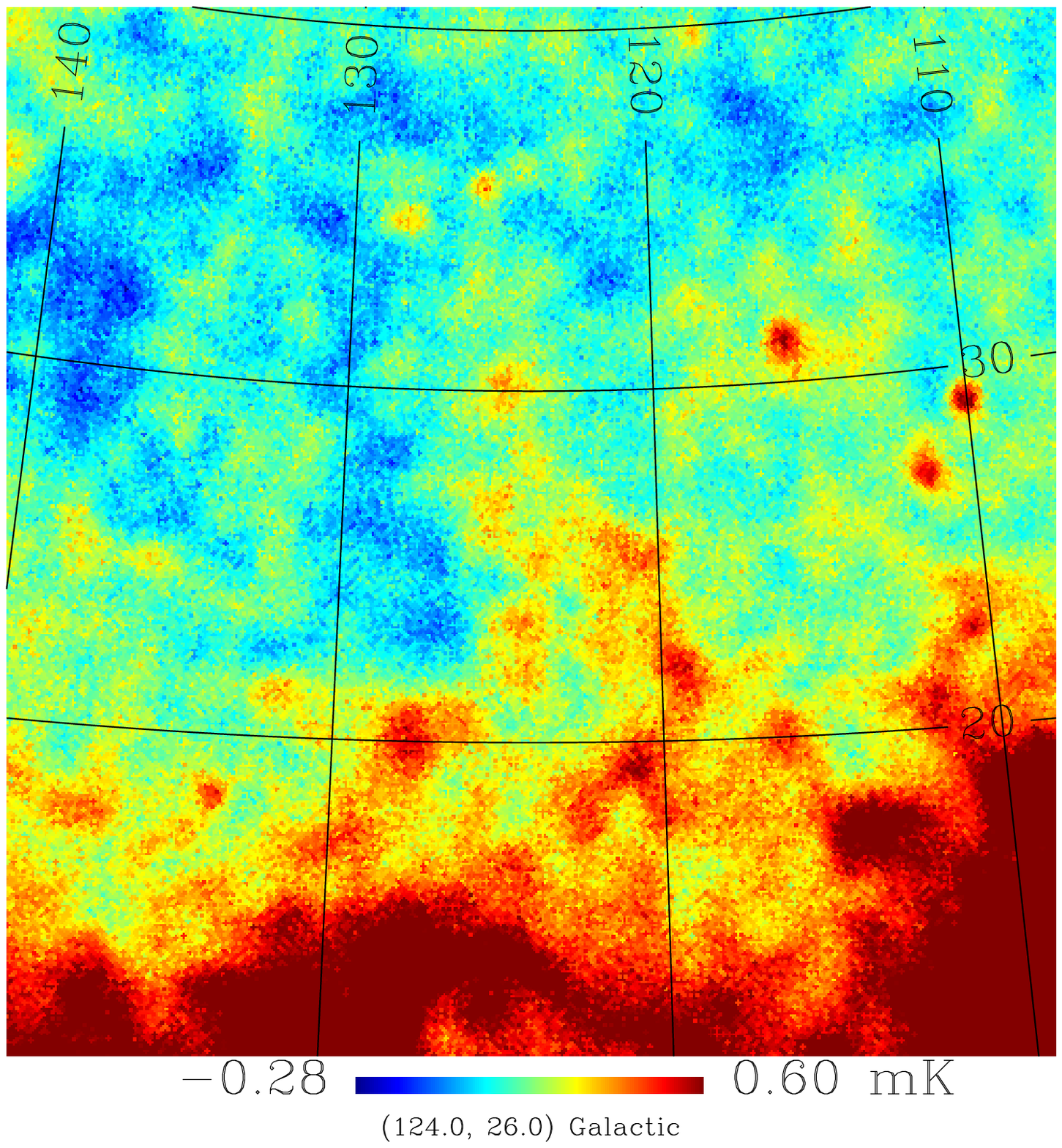}
\includegraphics[width=6cm]{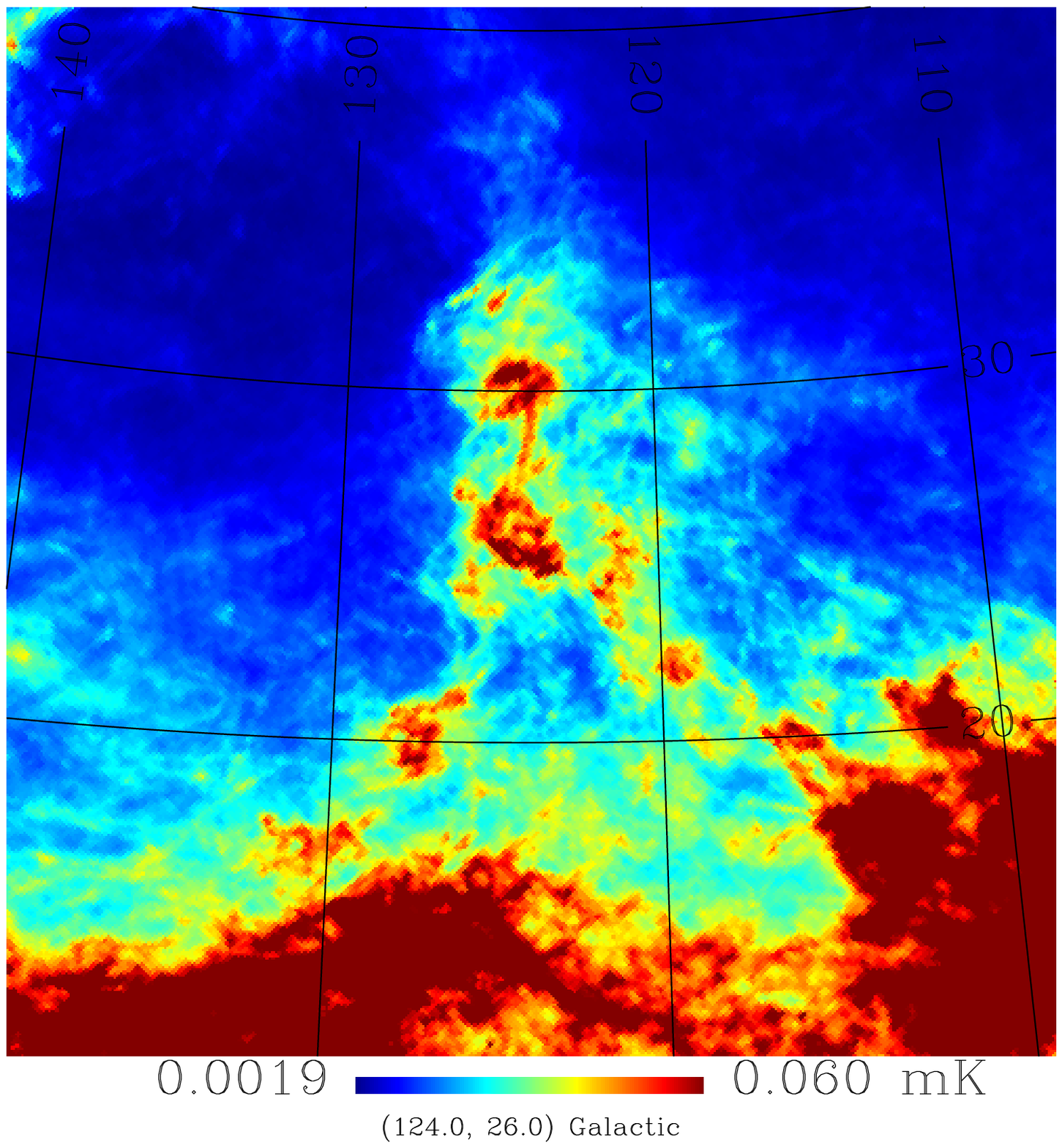}
\includegraphics[width=6cm]{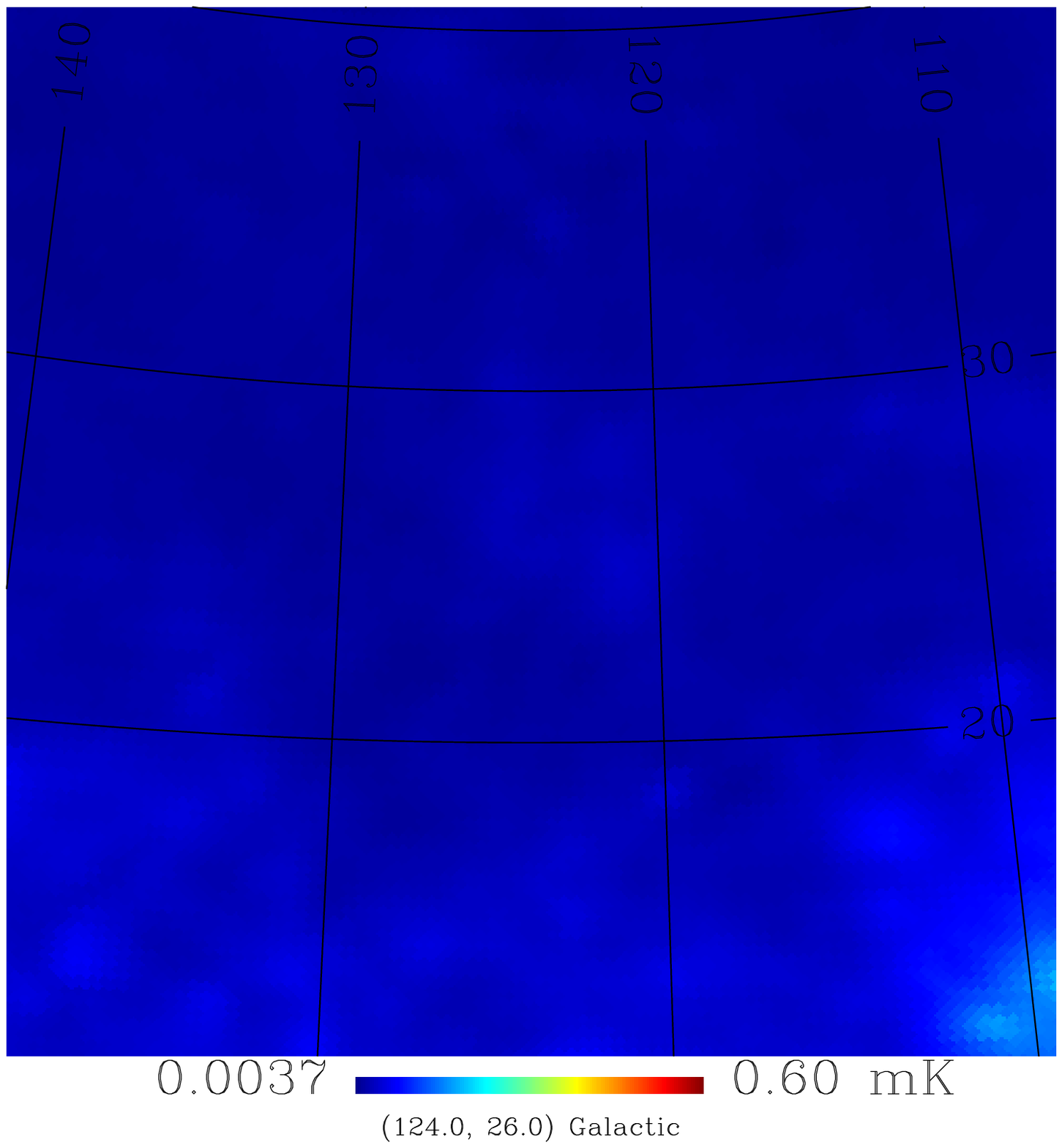}
\includegraphics[width=6cm]{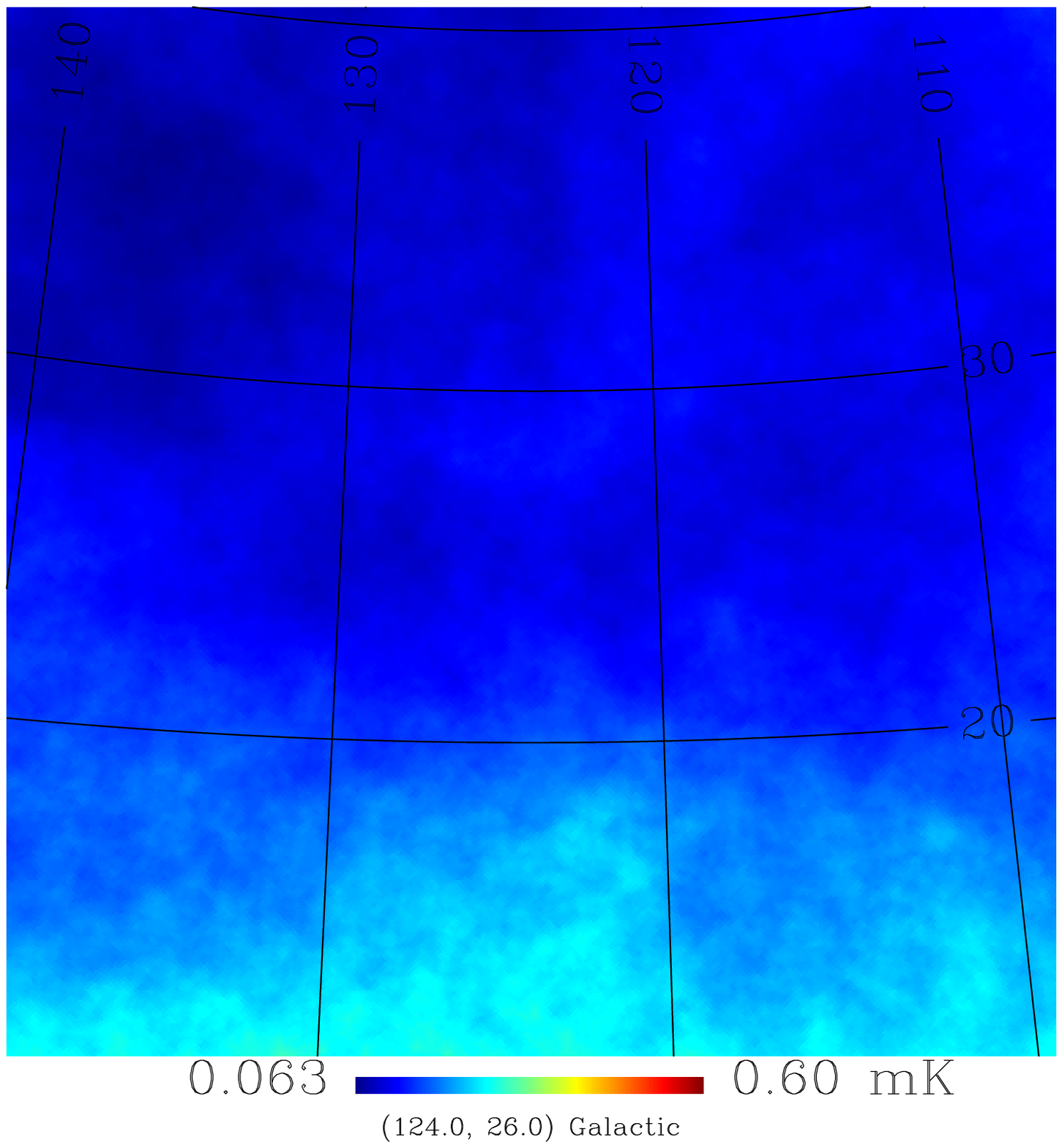}
\caption{Maps of the region of interest. Top: {\it WMAP} K band (left), 94\,GHz dust prediction (right,\cite{finkbeiner1999}). Bottom: free-free (left) and synchrotron (right) predictions at 23\,GHz based on the H$\alpha$ map (\cite{cliveff}) and the 408 MHz map (\cite{haslam}) assuming a spectral index $\beta_{\rm s}=-2.9$.}
\label{fig:gnom}
\end{center}
\end{figure}

\section{Description of the method}\label{sec:method}
In the following we describe the principles of operation of the harmonic-domain CCA; for further details we refer the reader to \cite{ricciardi2010}.
The CCA (\cite{bonaldi2006},\cite{ricciardi2010}) exploits second-order statistics to estimate the frequency scaling of the components from the statistics of data and noise.
For each position on the sky (each pixel) we write our data model as:
\begin{equation}
\mathbf{x}({r})=[\mathbf{B*H\,s}]({r})+\mathbf{n}({r}),\label{vect_m1}
\end{equation}
where  $\mathbf{{s}}$ is a vector whose elements contain the different components (CMB, free-free emission, synchrotron emission, thermal dust emission, AME); $\mathbf{{x}}$ and $\mathbf{{n}}$ are vectors too, each element containing respectively the data and the instrumental noise for each frequency.  The diagonal matrix $\mathbf{B}$ contains the instrumental beams for all frequency channels and $*$ denotes convolution. The matrix $\mathbf{H}$ is called the \emph{mixing matrix} and it contains the frequency scaling of the components for all the considered frequencies. 

By translating eq.~(\ref{vect_m1}) in the harmonic domain, the convolution becomes a multiplication and the data model becomes  a linear mixture. For each transformed mode ($\ell$) we can write:
\begin{equation}
\mathbf{X}(\ell)=\widetilde{\mathbf{B}}(\ell) \mathbf{HS(\ell)}+\mathbf{N}(\ell)\label{modhcca},
\end{equation}
where $\mathbf{X}$, $\mathbf{S}$, and $\mathbf{N}$ are the transforms
of $\mathbf{x}$, $\mathbf{{s}}$, and $\mathbf{n}$, respectively,
and $\widetilde{\mathbf{B}}$ is the transform of the matrix $\mathbf{B}$.
The cross-spectra of the data $\widetilde{\mathbf{C}}_{\mathbf{x}}(\ell)$, sources $\widetilde{\mathbf{C}}_{\mathbf{s}}(\ell)$ and noise, $\widetilde{\mathbf{C}}_{\mathbf{n}}(\ell)$, are related by:
\begin{equation}
\widetilde{\mathbf{C}}_{\mathbf{x}}(\ell)-\widetilde{\mathbf{C}}_{\mathbf{n}}(\ell)=\widetilde{\mathbf{B}}(\ell)\mathbf{H}\widetilde{\mathbf{C}}_{\mathbf{s}}(\ell)\mathbf{H}^T\widetilde{\mathbf{B}}^\dagger(\ell)
\label{fd_cca}
\end{equation}
where the dagger superscript denotes the adjoint matrix.  

The left-hand side of eq.~(\ref{fd_cca}) can be estimated from the data for a suitable set of spectral bins $\hat \ell$; this is used by CCA to estimate the mixing matrix and the source cross-spectra on the right-hand side of the equation. Due to the scaling ambiguity of the problem, the mixing matrix is normalized at a reference frequency.
To reduce the number of unknowns, the mixing matrix is parametrized through a parameter vector $\mathbf{p}$ (such that $\mathbf{H}=\mathbf{H}(\mathbf{p})$), by adopting suitable fitting relations for the spectra of the astrophysical components (as detailed in the next section). 

By using 2-dimensional discrete Fourier transforms it is possible to apply the harmonic-domain CCA to square sky patches. This approach is advisable as the mixing matrix varies on the sky. The HEALPix (\cite{gorski2005}) data on the sphere are projected on the plane tangential to the center of the patch and re-gridded with a suitable number of bins in order to correctly sample the original resolution. 
The patch size is obtained as a trade-off between the need of having uniform spectral properties of the foregrounds and enough statistics for a robust computation of the auto- and cross-spectra of the data. In the present analysis we use a patch size of  $30^\circ \times 30^\circ$. By using higher resolution data (e.g. {\it Planck} data) it is possible to reduce the patch size by carrying the analysis up to a larger multipole. 
We analyzed the sky patch centered on $(l,b)=(124^{\circ},26^{\circ})$ (shown in Fig.~\ref{fig:gnom}) and verified the stability of the results for a sample of patches shifted in latitude and longitude up to $10^{\circ}$.

\section{Description of the analysis}\label{sec:simulations}
We used the following datasets:
\begin{itemize}
\item {\it WMAP} 7-yr K, Ka, Q, V and W bands (\cite{jarosik2011}). All maps have been used at the original resolution except the K band, which has been smoothed to $1^\circ$ resolution to reduce the effect of beam asymmetry;
\item 408 MHz map (\cite{haslam}) to trace the synchrotron component;  
\item Predicted free-free emission at 23\,GHz based on the H$\alpha$ map by \cite{cliveff} corrected for dust absorption with the E(B-V) map by \cite{schlegel1998} assuming $f_{\rm d}=0.3$;
\item Predicted dust emission at 94\,GHz by \cite{finkbeiner1999}.
\end{itemize}
The different resolution of the data maps is accounted for with the beam matrix $\widetilde{\mathbf{B}}$ in eq.~({\ref{fd_cca}}). In particular, the beams are deconvolved for the mixing matrix estimation.
The noise properties for the {\it WMAP} maps have been computed by simulating different noise realizations starting from the input $N_{\rm obs}$ maps and $\sigma_0$ values. For the maps used as foreground templates (408\,MHz map, 23\,GHz free-free map and 94\,GHz dust prediction) we assumed a Gaussian noise at the 10\% level, which is much higher than the instrumental noise. This extra noise mimics the error on the template as a tracer of the true component, the 10\% level being indicative. We verified, however, that the results are not sensitive to the exact value assumed. 
We modeled the data as a mixture of five components: CMB, synchrotron emission, thermal dust emission, free-free emission and AME. Each frequency map contributes five elements to the mixing matrix, one for each component. The synchrotron, free-free and dust templates are used as frequency channels having only one mixing matrix entry, that of the corresponding component, different from zero. 

It is worth noting that, compared with previous work, we are not assuming correlation between the AME and thermal dust emission and we are not exploiting any template for the AME. Though the AME and the thermal dust emission are significantly correlated, the correlation is not supposed to be perfect, as, according to spinning dust models, the AME traces the distribution of smaller dust grains (PAHs). Indeed, the intensity ratio of AME and thermal dust is found to vary in the sky by  a factor of $\sim$2 (e.g., \cite{davies2006}). 

We estimated the frequency spectrum of the AME while assuming the spectra of the other components to be known. 
For CMB we used the usual black-body law with temperature of 2.726\,K (\cite{fixsen1997}); for free-free we used the model $T_{\rm RJ,ff}(\nu)\propto G \times (\nu/10)^{-2}$ where $G=3.96 (T_4)^{0.21}(\nu/40)^{-0.14}$ is the Gaunt factor, which is responsible for the departure from a pure power-law behavior, and $T_4=0.7$ is the electron temperature $T_{\rm e}$ in units of $10^4$\,K. The thermal dust emission has been modeled as a grey-body with temperature $T_{\rm d}=18$\,K and spectral index $\beta_{\rm d}=1.7$. This is consistent with the 1-component dust model by \cite{finkbeiner1999}; {\it Planck} data, extending the observations up to 857\,GHz, will allow a refinement of this model. For synchrotron we assumed a power-law scaling with spectral index $\beta_{\rm s}=-2.9$. 

\begin{figure}
\begin{center}
\includegraphics[width=11cm]{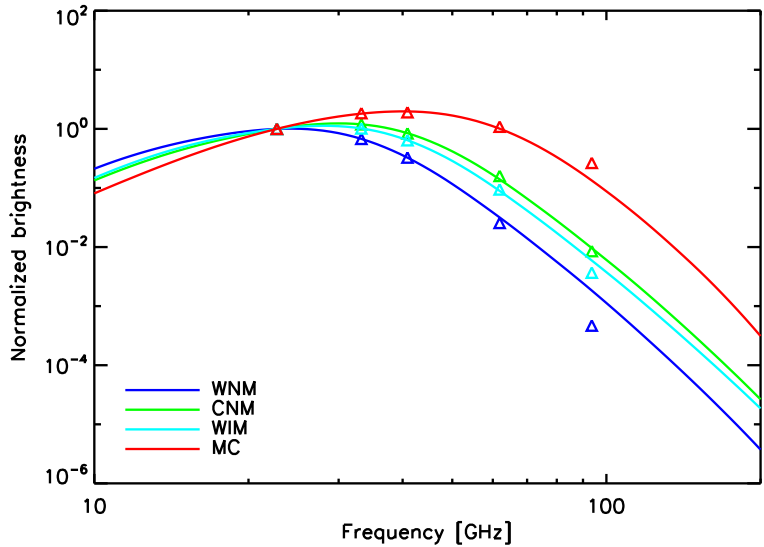}
\caption{Theoretical spinning dust models produced with SpDust (solid lines) and fitted with CCA parametric model (triangles, same color of the model). Input SpDust parameters and best-fit parameters for the CCA model are provided in Table~\ref{tab:commontab}.}
\label{fig:ametest}
\end{center}
\end{figure}

For the AME we used the parametric relation proposed by \cite{bonaldi2007}: 
\begin{equation}
\log T_{\rm RJ,ame}(\nu)  \propto  \left(\frac{m_{60}\log
\nu_{\rm p}}{\log(\nu_{\rm p}/60)}+2\right)\log \nu + \frac{m_{60}(\log\nu)^2}{2\log(\nu_{\rm p}/{60})} \label{tspin},
\end{equation}
which is a parabola in the $\log (\nu)-\log (S)$ plane parametrized in terms of peak frequency $\nu_{\rm p}$ and slope at 60\,GHz $m_{60}$. The CCA solves for the two parameters $m_{60}$ and $\nu_{\rm p}$, so the accuracy of the estimation of the AME peak frequency $\nu_{\rm p}$ is related to the ability of the parametric relation to reproduce the true spectrum. We verified, however, that this parametric model can fit a wide range of spinning dust spectra. This is shown in Fig.~\ref{fig:ametest},  comparing spinning dust spectra produced with the SpDust (\cite{spdust1}, \cite{spdust2}) code with the results of the fit of eq.~(\ref{tspin}) obtained by minimizing the $\chi^2$ for the set of frequencies considered in this work. The physical models that we consider are: warm neutral medium (WNM), cold neutral medium (CNM), warm ionised medium (WIM), and molecular cloud (MC).    
In general, the fits are accurate up to 61\,GHz, while at 94\,GHz the parametric relation may not be able to reproduce the input spectra in detail. This is a consequence of fitting complex spectra with only a few parameters, the fit being less accurate where the AME is weaker. 
The best-fit parameters that we obtain, reported in Table~\ref{tab:commontab}, vary significantly from one input model to another.
\begin{table}
\begin{center}
\caption{Parameters for the spectra in Fig.~\ref{fig:ametest}. The SpDust input parameters are: the total hydrogen number density $n_{\rm H}$, the gas temperature $T$, the intensity of the radiation field relative to the average interstellar radiation field $\chi$, the hydrogen ionization fraction $x_{\rm H}=n_{\rm H^{+}} / n_{\rm H}$ and the ionized carbon fractional abundance $x_{\rm C}=n_{\rm C^{+}} / n_{\rm H}$. We also report the CCA spectral parameters $\nu_{\rm p}$, $m_{60}$ [eq.~(\ref{tspin})].}
\begin{tabular}{llllllll}
\hline
model name&$n_{\rm H}$\,[cm$^{-3}$]&T\,[K]&$\chi$&$x_{\rm H}$&$x_{\rm C}$&$\nu_{\rm p}$&$m_{60}$\\
\hline
WNM &0.4&6000 &1.00 &0.10  &0.0003 & 24.34&7.64  \\
CNM &30.0&100 &1.00 &0.0012&0.0003 & 29.26&5.31  \\
WIM &0.1&8000 &1.00 &0.99  &0.001  & 27.64&5.99  \\
MC  &300  &20 &0.01 &0.0   &0.0001 & 38.13&2.23  \\
\hline
\end{tabular}
\label{tab:commontab}
\end{center}
\end{table}
\section{Test on simulated data}\label{sec:simul}
We simulated {\it WMAP} 7-yr data by assuming monochromatic band-passes centered at the central frequency of the realistic bands, Gaussian beams at the nominal values, Gaussian noise generated according to realistic (spatially varying) RMS. Our model of the sky consists of the following components:
\begin{itemize}
\item CMB emission constrained by the best-fit power spectrum model to {\it WMAP} 7-yr;
\item Synchrotron emission given by \cite{haslam} scaled in frequency with a power-law model with a spatially varying synchrotron spectral index $\beta_{\rm s}$ as modeled by \cite{giardino};
\item Free-free component modeled by the H$\alpha$ map of \cite{cliveff} corrected for dust absorption with the E(B-V) map from \cite{schlegel1998} with $f_{\rm d}=0.3$ and scaled in frequency with a power-law with fixed spectral index of $-2.14$;
\item Thermal dust emission modeled with the 100\,$\mu$m map from \cite{schlegel1998} scaled in frequency with the best-fit model by \cite{finkbeiner1999} which consists of two grey-body laws having different temperatures and emissivity indices; 
\item AME emission modeled by the E(B$-$V) map from \cite{schlegel1998} with intensity at 23\,GHz calibrated on the results by \cite{davies2006} for average intermediate-latitude conditions and scaled according to SpDust models. 
\end{itemize} 
It is worth noting that this sky model is more complex than the spectral model assumed in the component separation. The synchrotron spectral index is spatially varying and both the thermal dust AME spectra, though spatially constant, are not generated with the CCA parametric models. This has been done purposely, to reflect a more realistic situation.
To test the ability of our pipeline to recover the AME spectrum in different situations, we adopted two SpDust models for the AME: one peaking at 26\,GHz and the other peaking at 19\,GHz. 
\begin{figure}
\begin{center}
\includegraphics[width=11cm]{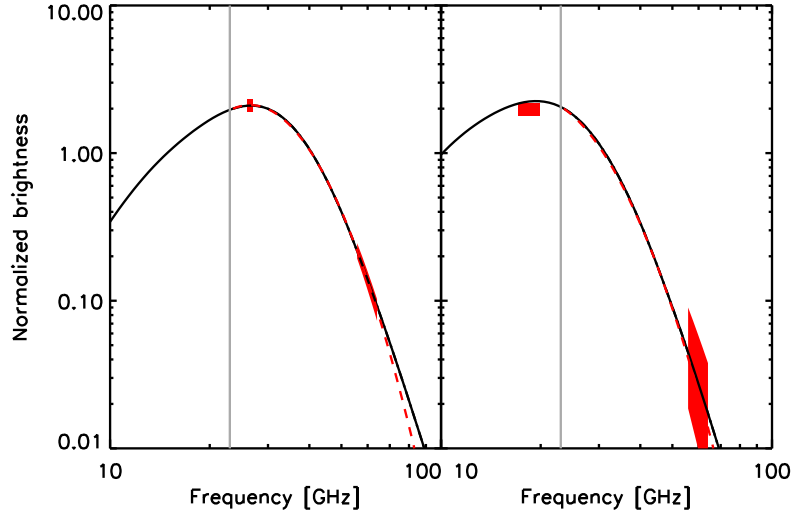}
\caption{Comparison of true input AME spectra (black solid lines) with CCA estimates (red dashed lines) for two AME models, one peaking at 26\,GHz (left) and one at 19\,GHz (right). The red shaded areas show the recovered $\nu_{\rm p}$ and $m_{60}$ with 1$\sigma$ error bar. The results for the peak frequency are $\nu_{\rm p}=26.5 \pm 0.5$ and $\nu_{\rm p}=18.5 \pm 1.5$ for the left and right panels respectively. The grey vertical lines indicate the frequency of 23\,GHz which sets the limit of our data.}
\label{fig:amespec_sim}
\end{center}
\end{figure}

We performed an estimate of the AME spectrum on the simulated data following the same procedure applied to the real data. The results are presented in Fig.~\ref{fig:amespec_sim}. For AME peaking at 26\,GHz the recovery of the spectrum is very accurate; the result on the peak frequency is $\nu_{\rm p}=26.5 \pm 0.5$. The accuracy of the estimation is helped by the weak synchrotron and free-free contamination in this region of the sky (in a more general case the error on $\nu_{\rm p}$ would be of a few GHz). With the AME peaking at 19\,GHz the errors increase, but the recovery of the spectrum is still satisfactory; we obtain $\nu_{\rm p}=18.5 \pm 1.5$. Our pipeline is able to distinguish very clearly between the two input models and to correctly estimate the peak frequency of the AME in both cases. 

\section{Results}\label{sec:real}
\begin{figure}
\begin{center}
\includegraphics[width=11cm]{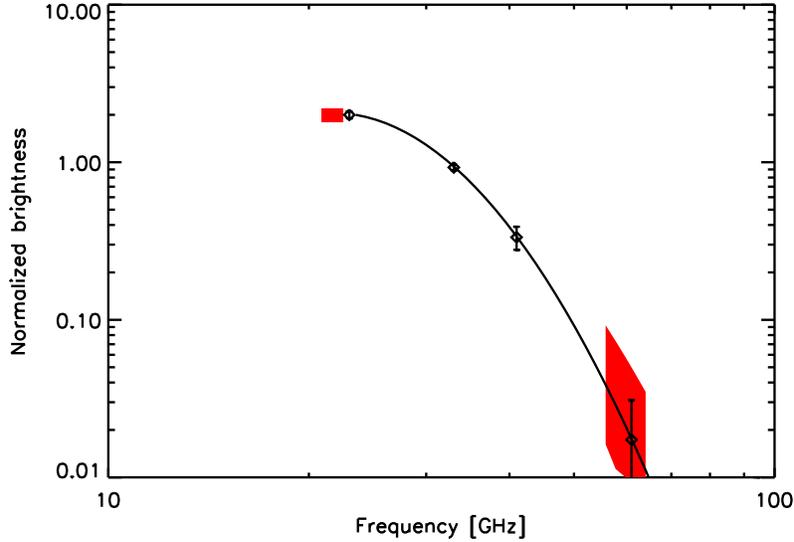}
\caption{Best-fit AME model (black line) and results in terms of $\nu_{\rm p}$ and $m_{60}$ considering 1$\sigma$ uncertainties (red areas). The peak frequency results $\nu_{\rm p}= 21.7\pm 0.8$. Diamonds with error bars are the results in terms of normalized intensity at {\it WMAP} bands and related uncertainties.}
\label{fig:amespec_real}
\end{center}
\end{figure}

The results for the AME spectrum on the NCP are showed in Fig.~\ref{fig:amespec_real}; the estimated peak frequency is $\nu_{\rm p}=21.7\pm0.8$. 
The diamonds with error bars are the AME mixing matrix elements at {\it WMAP} frequencies and related uncertainties derived from the results on $\nu_{\rm p}$ and $m_{60}$. 

By comparing these data points with theoretical models produced with SpDust we can constrain the physical properties of the medium with the hypothesis of spinning dust emission. Some relevant parameters of the SpDust code are: hydrogen number density $n_{\rm H}$ (cm$^{-3}$); gas temperature $T$ (K); intensity of the radiation field with respect to average Galactic conditions, $\chi$, and hydrogen ionization fraction, $x_{\rm H}$. Advanced parameters describe the dust grains and define the abundance of other elements.  

The hydrogen density $n_{\rm H}$ is sensitive to the position of the peak and in general to the shape of the spectrum in the considered frequency range. It is also degenerate with other parameters, in particular the gas temperature $T$. We investigated this degeneracy by sampling the 2-dimensional $n_{\rm H}$--$T$ space with a grid approach while conditioning the remaining parameters. For each point in the parameter space we computed the likelihood as $\Lambda=\exp(-\chi^2/2)$ where $\chi^2$ is the standard chi-square. In computing the $\chi^2$ we did not consider the 94\,GHz data point, which could be biased, as verified through simulations (see Sect.~\ref{sec:simul}). We repeated the likelihood estimation for different choices of the parameters $x_{\rm H}$ and $\chi$ (varying all those parameters simultaneously would be computationally too demanding). As the CCA outputs are in normalized units, we compared both the model and the data after scaling each of them by the sum of the intensities at 23, 33, 41 and 61\,GHz. 
\begin{figure}
\begin{center}
\includegraphics[width=7.5cm]{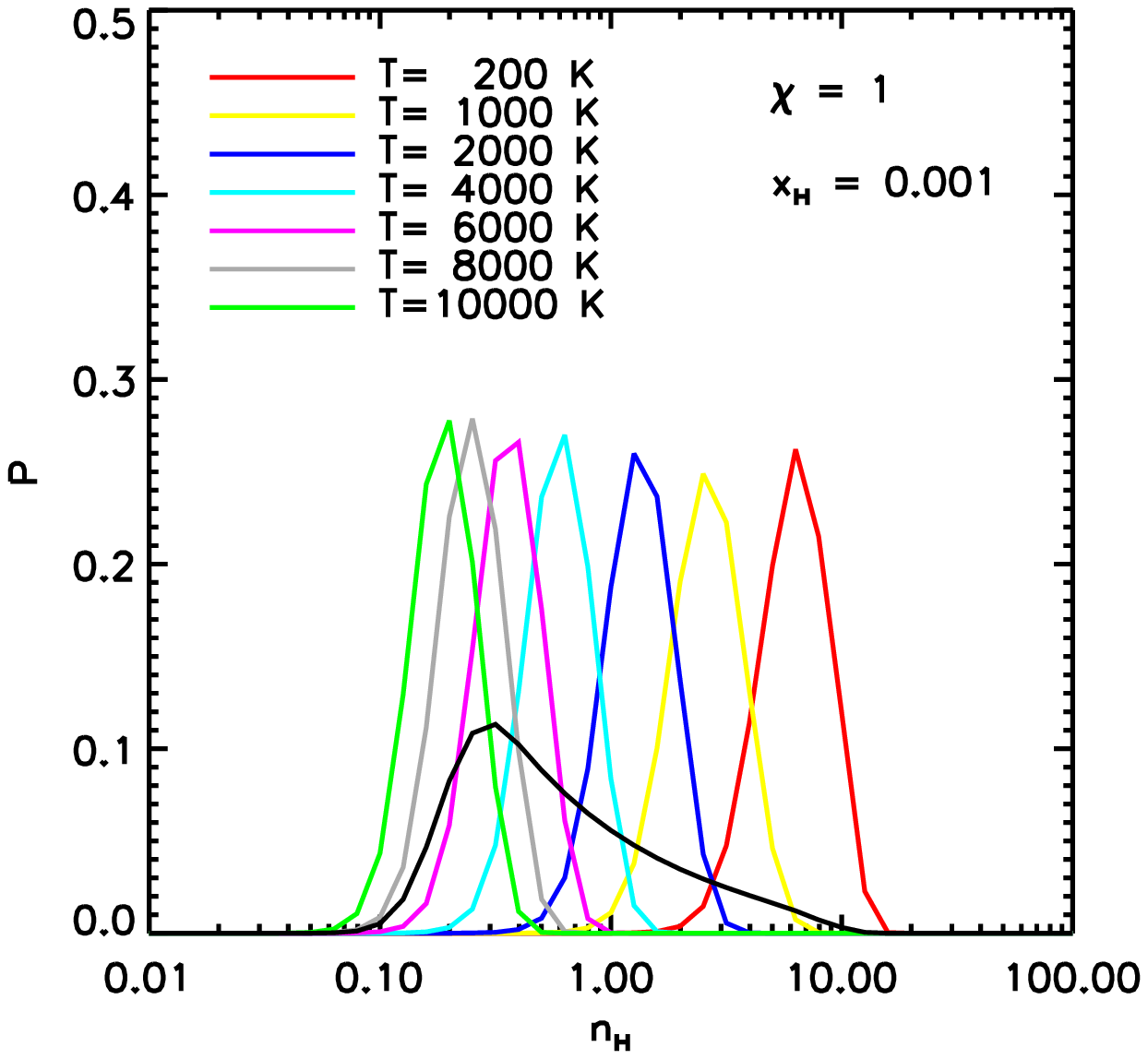}
\includegraphics[width=7.5cm]{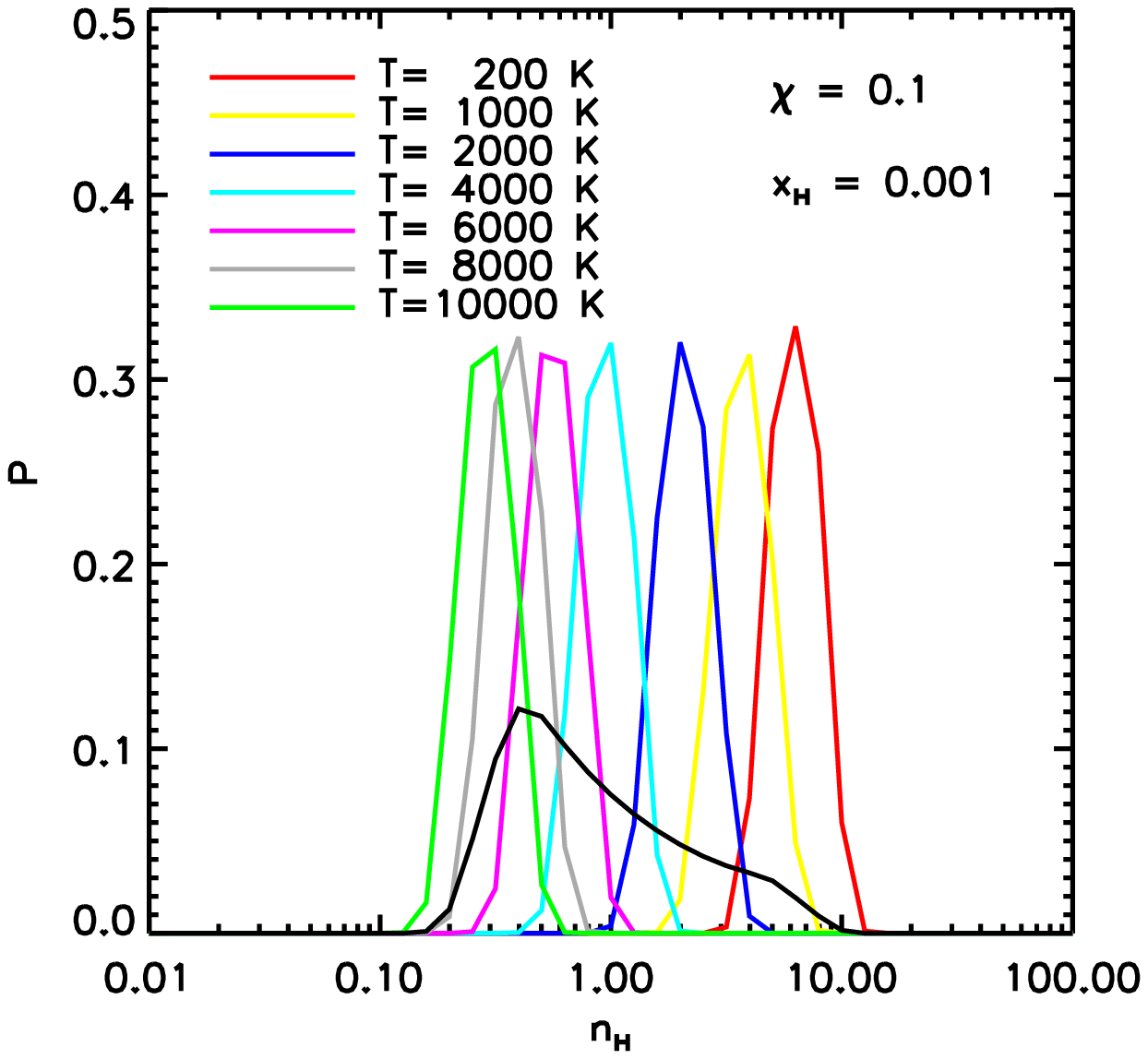}
\includegraphics[width=7.5cm]{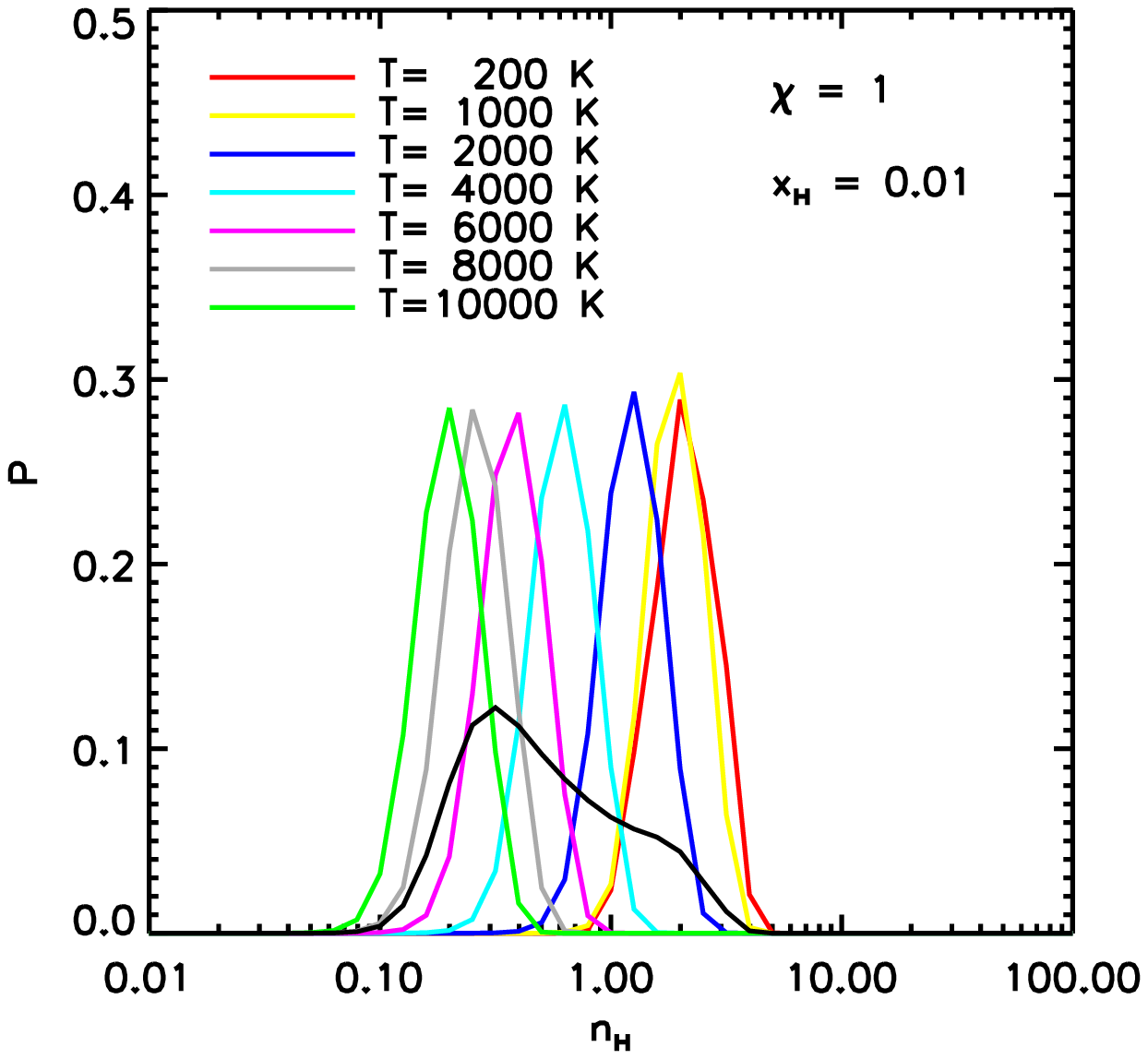}
\includegraphics[width=7.5cm]{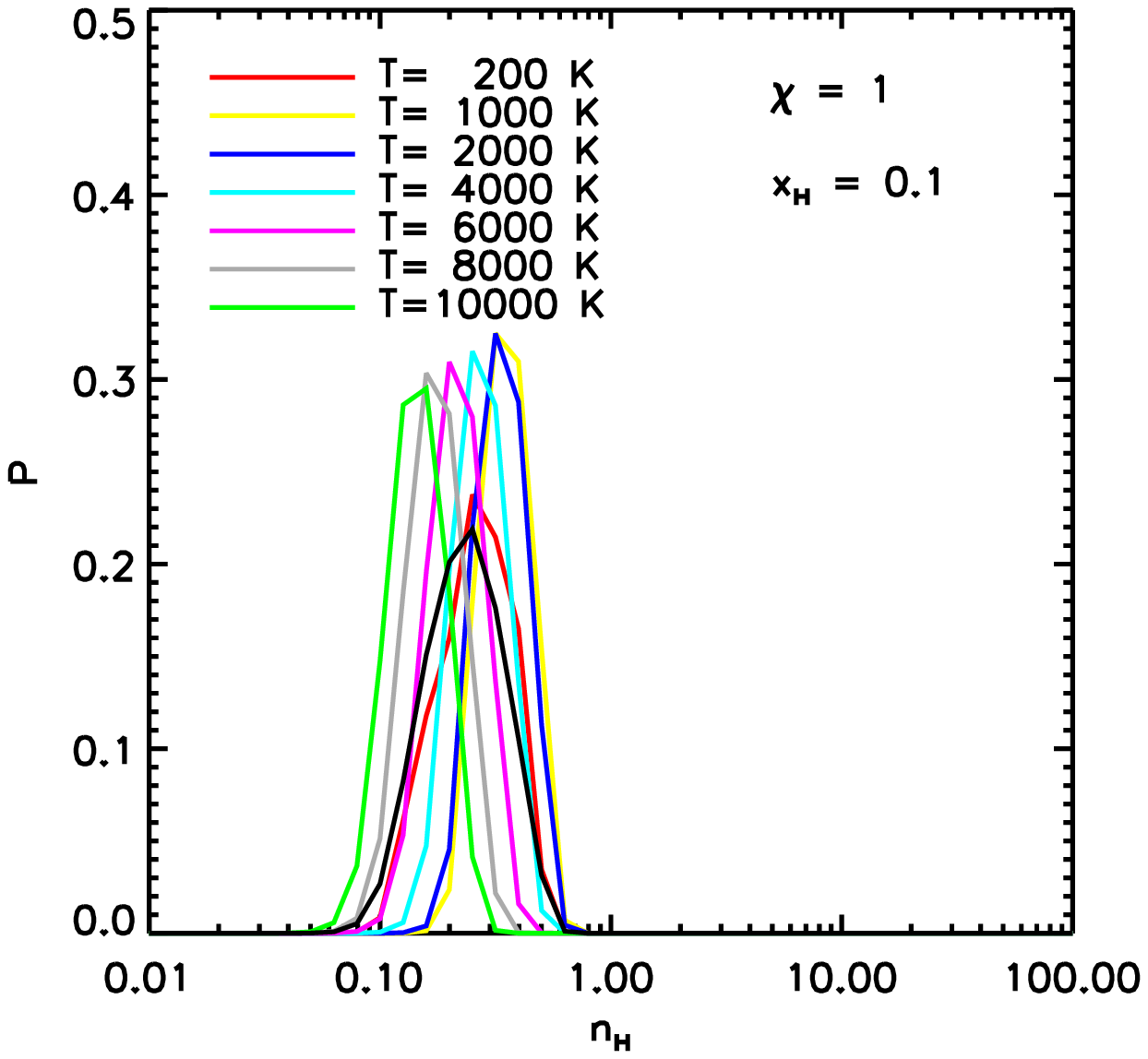}
\caption{Colored lines: likelihoods for $n_{\rm H}$ for different values of $T$ as detailed in the legend; black line: likelihood for $n_{\rm H}$ marginalized over $T$. Top left: $x_{\rm H}=0.001$, $\chi=1$; top right: same of top left with $\chi=0.1$; bottom: same as top left with $x_{\rm H}=0.01$ (left) and $x_{\rm H}=0.1$ (right).}
\label{fig:like1}
\end{center}
\end{figure}
\begin{figure}
\begin{center}
\includegraphics[width=11cm]{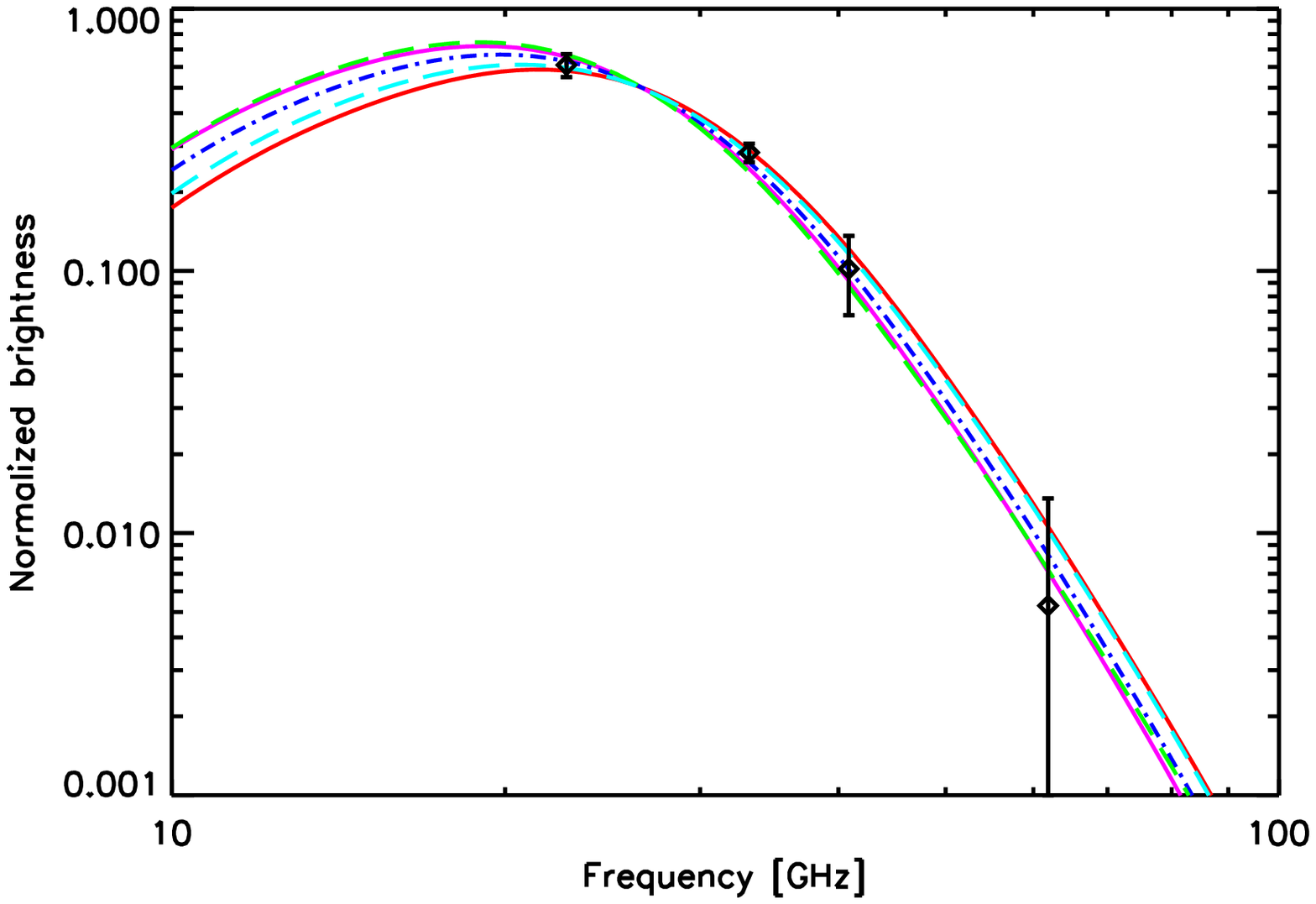}
\caption{Comparison between data (diamonds and error bars) and different SpDust models. Red: $n_{\rm H}=0.25$\,cm$^{-3}$ $T=6000$\,K, $x_{\rm H}=0.1$, $\chi=1$; magenta: same as red with $x_{\rm H}=0.01$; blue: $n_{\rm H}=0.30$\,cm$^{-3}$ $T=6000$\,K, $x_{\rm H}=0.01$, $\chi=1$; green: $n_{\rm H}=0.40$\,cm$^{-3}$ $T=6000$\,K, $x_{\rm H}=0.001$, $\chi=1$; cyan: same as blue with $\chi=0.1$.}
\label{fig:fig_fits}
\end{center}
\end{figure}

The results of the $n_{\rm H}$--$T$ parameter estimation are shown in Fig.~\ref{fig:like1}. The colored lines show the likelihoods for $n_{\rm H}$ for different values of $T$ as detailed in the legend, while the black lines show the likelihood for $n_{\rm H}$  marginalized over $T$ in the range 100--10000\,K. Different panels correspond to different choices for the other parameters: $x_{\rm H}=0.001$, $\chi=1$, on the top left; $x_{\rm H}=0.001$, $\chi=0.1$, on the top right, $x_{\rm H}=0.01$, $\chi=1$, on the bottom left and $x_{\rm H}=0.1$, $\chi=1$ on the bottom right. The marginalized statistics for $n_{\rm H}$ are reported in Table \ref{tab:stats} ($\Delta n_{\rm H}^-$ and $\Delta n_{\rm H}^+$ being the lower and upper 1$\sigma$ error on $n_{\rm H}$). The $n_{\rm H}$--$T$ degeneracy weakens the constraint on $n_{\rm H}$ in particular on the high-density side of the distribution. The degeneracy decreases with increasing hydrogen ionization fraction. 

The best results are obtained for densities of 0.2--0.4\,cm$^{-3}$, typical of WIM/WNM conditions. In Fig.~\ref{fig:fig_fits} we show the data compared to some SpDust models to exemplify the quality of the fit. The comparison of the red ($n_{\rm H}=0.25$\,cm$^{-3}$, $T=6000$\,K, $x_{\rm H}=0.1$, $\chi=1$) and magenta ($n_{\rm H}=0.25$\,cm$^{-3}$, $T=6000$\,K, $x_{\rm H}=0.01$, $\chi=1$) lines illustrates the effect of $x_{\rm H}$.  The blue line corresponds to $n_{\rm H}=0.3$\,cm$^{-3}$, $T=6000$\,K, $x_{\rm H}=0.31$, $\chi=1$; compared to the magenta line, it illustrates the effect of increasing $n_{\rm H}$. Finally, the cyan ($n_{\rm H}=0.40$\,cm$^{-3}$, $T=6000$\,K, $x_{\rm H}=0.001$, $\chi=0.1$) and green ($n_{\rm H}=0.40$\,cm$^{-3}$, $T=6000$\,K, $x_{\rm H}=0.001$, $\chi=1$) lines show the effect of $\chi$. The availability of data at frequencies of a few GHz will be very valuable to improve the constraints on such parameters. 
 
To complement the physical description of the medium we computed the hydrogen column density $N_{\rm H}$ (cm$^{-2}$) from the brightness at 100\,$\mu$m, $I_{100}$, through  $I_{100}$\,[MJy\,sr$^{-1}] = (0.69 \pm 0.03) \times N_{\rm H}\,[10^{20}\,$cm$^{-2}$] (\cite{boulanger1996}). We used the IRIS (\cite{IRIS}) band 4 (100\,$\mu$m) data and integrated the emission inside a $1^\circ$ beam centered in $l=25^{\circ}$, $b=125^{\circ}$, corresponding to bright dust emission (see Fig.~\ref{fig:gnom}), to compute a representative value for $I_{100}$. This analysis yields $N_{\rm H}=(2.9 \pm 0.2)\times 10^{20}$\,cm$^{-2}$, which is a plausible value for the diffuse ISM environment at intermediate latitudes.

\begin{table}
\begin{center}
\caption{Statistics for $n_{\rm H}$ marginalized over $T$ corresponding to the results of Fig.\ref{fig:like1}.}
\begin{tabular}{lllll}
\hline
$x_{\rm H}$&$\chi$&$n_{\rm H}$&$\Delta n_{\rm H}^-$&$\Delta n_{\rm H}^+$\\
\hline
0.001&1&0.3&0.1&0.9 \\
0.01&1&0.3&0.1&0.7  \\
0.1&1&0.25&0.09&0.07   \\
0.001&0.1&0.4&0.1&1.2\\
\hline
\end{tabular}
\label{tab:stats}
\end{center}
\end{table}

\section{Conclusions} \label{sec:conclu}
We have studied the spectrum of the diffuse AME with {\it WMAP} 7-yr and ancillary data in the North Celestial Pole (NCP) region of the sky. In this region the AME dominates the low-frequency emission, as both synchrotron and free-free are faint. Previous template-fitting analysis by \cite{davies2006} found that the AME spectrum in this region is consistent with a power-law; the same would apply to most diffuse AME at intermediate latitudes (\cite{banday2003}, \cite{davies2006}, \cite{bonaldi2007}, \cite{dobler2008}, \cite{mamd}, \cite{ghosh2011}). This favors a low peak frequency ($\nu_{\rm p}<23$\,GHz).

For our analysis we rely on the CCA component separation method, which exploits the data auto- and cross-spectra to estimate the frequency spectra of the components in terms of a set of spectral parameters. Our method models the AME as a peaked spectrum and fits for the peak frequency, $\nu_{\rm p}$, and the slope at 60\,GHz, $m_{60}$. We verified with simulations that we are able to correctly recover the AME spectrum and more specifically the peak frequency, even when it is below 23\,GHz. We get $\nu_{\rm p}=21.7\pm0.8$\,GHz, which is both a confirmation and an improvement of the previous results. This result relies on the assumption that our parametric model for the AME is a good representation of the true spectrum, which has been verified for a wide range of theoretical spinning dust models.

Using the SpDust code, we linked the estimated spectrum to the local physical conditions with the hypothesis of spinning dust emission. We investigated in particular the hydrogen density ($n_{\rm H}$\,[cm$^{-3}$]), which is sensitive to the position of the peak, and its degeneracy with the gas temperature $T$\,[K]. The densities that we get are those typical of WIM/WNM conditions (0.2--0.4\,cm$^{-3}$). For a low hydrogen ionization fraction ($n_{\rm H} \leq10^{-2}$), densities up to a few cm$^{-3}$ are allowed by the $n_{\rm H}$--$T$ degeneracy. Lower radiation fields ($\chi$) require slightly lower $n_{\rm H}$. 
By considering 100\,$\mu$m data for our region we obtain a hydrogen column density $N_{\rm H} \sim10^{20}$\,cm$^{-2}$. 
Overall, the recovered AME spectrum is found to be consistent with that predicted by spinning dust models for plausible physical conditions. 

\section{Acknowledgements}
S. Ricciardi acknowledges support by MIUR through PRIN 2009 grant n. 2009XZ54H2





\bibliographystyle{abbrv}
\bibliography{biblio}

\end{document}